\documentclass[11pt,a4paper]{article}
\usepackage{jheppub_kim}

\usepackage{subfigure,amsmath,amssymb ,amsfonts, latexsym}

\usepackage[notcite,color,final
                              ]{showkeys}
\definecolor{refkey}{gray}{.25}
\definecolor{labelkey}{gray}{.25}

\newcommand{\ud}{\mathrm{d}}
\def \pd {\partial}

\newcommand{\be}{\begin{equation}}
\newcommand{\ee}{\end{equation}}
\newcommand{\beqa}{\begin{subequations}\begin{eqnarray}}
\newcommand{\eeqa}{\end{eqnarray}\end{subequations}}

\usepackage{color}

\begin{document}

\title{Phantom crossing  and quintessence
limit in extended nonlinear massive gravity}

\author[a,b,c]{Emmanuel N. Saridakis}

\affiliation[a]{Physics Division, National Technical University of Athens,
15780 Zografou Campus, Athens, Greece}

\affiliation[b]{CASPER, Physics Department, Baylor University,
Waco, TX  76798-7310, USA}

\affiliation[c]{Institut d'Astrophysique de Paris, UMR 7095-CNRS,
Universit\'e Pierre \& Marie Curie, 98bis boulevard Arago, 75014 Paris,
France}

\emailAdd{Emmanuel$_-$Saridakis@baylor.edu}




\abstract{
We investigate the cosmological evolution in a universe governed by the
extended, varying-mass, nonlinear massive gravity, in which the graviton
mass is promoted to a scalar-field. We find that the dynamics, both in
flat and open universe, can lead the varying graviton mass to zero at late
times, offering a natural
explanation for its hugely-constrained observed value. Despite the limit of
the scenario towards standard quintessence, at early and intermediate times
it gives rise to an effective dark energy sector of a dynamical nature,
which can also lie in the phantom regime, from which it always exits
naturally, escaping a Big-Rip. Interestingly enough, although the
motivation of massive gravity is
to obtain an IR modification, its varying-mass extension  in cosmological
frameworks leads to early and intermediate times modification instead.
}


\maketitle

\section{Introduction}

The idea of adding mass to the graviton is quite old \cite{Fierz:1939ix},
but the necessary nonlinear terms \cite{Vainshtein:1972sx} that can give
rise to continuity of the observables  \cite{vdam,vdam2} lead also to
Boulware-Deser (BD) ghosts \cite{Boulware:1973my}, making the theory
unstable. However, recently, a nonlinear extension of   massive gravity
has been constructed \cite{deRham:2010ik, deRham:2010kj} such that   
the 
Boulware-Deser ghost  is systematically removed (see
\cite{Hinterbichler:2011tt} for a review). The theoretical and
phenomenological advantages, amongst which is the universe
self-acceleration arising exactly from this IR gravity modification,
brought this theory to a significant attention
\cite{Koyama:2011yg,Hassan:2011hr,deRham:2011rn,CuadrosMelgar:2011yw,
D'Amico:2011jj,Hassan:2011zd,Kluson:2011qe,Gumrukcuoglu:2011ew,
Volkov:2011an,vonStrauss:2011mq,Comelli:2011zm,Hassan:2011ea,
Berezhiani:2011mt,Gumrukcuoglu:2011zh,Khosravi:2011zi,Brihaye:2011aa,
Buchbinder:2012wb,Ahmedov:2012di,Bergshoeff:2012ud,Crisostomi:2012db,
Paulos:2012xe,Hassan:2012qv,Comelli:2012vz,Sbisa:2012zk,Kluson:2012wf,
Tasinato:2012mf,Morand:2012vx,Cardone:2012qq,Baccetti:2012bk,Gratia:2012wt,
Volkov:2012cf,DeFelice:2012mx,Gumrukcuoglu:2012aa,deRham:2012kf,
Berg:2012kn,D'Amico:2012pi,Baccetti:2012re,Fasiello:2012rw,D'Amico:2012zv,
Baccetti:2012ge,Cai:2012db,Langlois:2012hk,deRham:2012ew}.

Despite the successes of massive gravity, in the case where the physical
and the fiducial metrics have simple homogeneous and isotropic forms the
theory proves to be unstable at the perturbation level
\cite{DeFelice:2012mx}, which led some authors to start constructing less
symmetric models  
\cite{D'Amico:2011jj,Gumrukcuoglu:2012aa}. However, in 
\cite{Huang:2012pe} a different approach was followed, that is expected to
be free of the above instabilities, namely to extend
the theory in a way that the graviton mass is varying, and this was
achieved by introducing an extra scalar field which coupling to the
graviton potentials produces an effective, varying, graviton mass.
 
In this work we desire to explore the cosmological implications of this
``extended'', varying-mass, massive gravity, in both flat and open
universe. As we show, at least in simple
cosmological ansatzes, the dynamics leads the varying graviton mass
to zero, or to a suitably chosen very small value in agreement with
observations, at late times, and thus the theory has as a limit the
standard
quintessence paradigm. However, at intermediate times the varying graviton
mass leads to very interesting behavior, with a dynamical effective dark
energy sector which can easily lie in the phantom regime. Strictly
speaking, although the motivation of massive gravity is to obtain an IR
modification, its extension  in cosmological frameworks  leads rather to 
early and intermediate times modification, and thus to a radical UV
modification instead.

\section{Extended nonlinear massive gravity}
\label{model}

Let us briefly review the ``mass-varying massive gravity'' that was
recently presented in  \cite{Huang:2012pe}. Their construction
is based on the promotion of the graviton mass to a scalar-field function
(potential), with the additional insertion in the action of this scalar
field's kinetic term and standard potential. Since such a modification is
deeper than allowing for a varying mass, we prefer to call it ``extended''
nonlinear massive gravity.

In such a construction the action writes as
 \begin{eqnarray} \label{action0}
&& S= \int d^4x \sqrt{-g} \left[ \frac{M_P^2}{2} R + V(\psi) ( U_2 +
\alpha_3 U_3 + \alpha_4 U_4)
-  \frac{1}{2} \pd_\mu \psi \pd^\mu
\psi -W(\psi)
\right],
\end{eqnarray}
where $M_p$ is the Planck mass, $R$ the Ricci scalar, and $\psi$ is the new
scalar field with $W(\psi)$ its standard potential and $V(\psi)$ its
coupling potential which spontaneously breaks general covariance.
Furthermore, as usual $\alpha_3$ and $\alpha_4$ are dimensionless
parameters, and the  graviton potentials are given by
\begin{align}
U_2  =  \mathcal{K}^\mu_{[\mu}\mathcal{K}^\nu_{\nu]} , \quad \quad
U_3  = \mathcal{K}^\mu_{[\mu}\mathcal{K}^\nu_{\nu}\mathcal{K}^\rho_{\rho]} 
, \quad \quad
U_4  =
\mathcal{K}^\mu_{[\mu}\mathcal{K}^\nu_{\nu}\mathcal{K}^\rho_{\rho}\mathcal{
K}^\sigma_{\sigma]} ,
\end{align}
with $\mathcal{K}^\mu_{[\mu} \mathcal{K}^\nu_{\nu]}
=(\mathcal{K}^\mu_{\mu} \mathcal{K}^\nu_{\nu}
-\mathcal{K}^\mu_{\nu}\mathcal{K}^\nu_{\mu})/2$ and similarly for the
other antisymmetric expressions, and
\be
\mathcal{K}^\mu_\nu = \delta^\mu_\nu-\sqrt{g^{\mu\rho}f_{AB}\pd_\rho \phi^A
\pd_\nu \phi^B }   .
\ee
As in standard massive gravity
$f_{AB}$ is a fiducial metric, and the four $\phi^A(x)$ are the
St\"{u}ckelberg scalars introduced to restore general covariance
\cite{ArkaniHamed:2002sp}, and in the particular case where the $f_{AB}$
is the Minkowski metric they form Lorentz
4-vectors in the internal space and the theory   presents a global 
Poincar\'e symmetry, too. Finally, one can show that the above extended
massive
gravity is still free of the the Boulware-Deser ghost \cite{Huang:2012pe}.

\section{Cosmological equations}
   \label{cosmequations}

Let us now examine cosmological scenarios in a universe governed by the
extended nonlinear massive gravity. Firstly, in order to obtain a realistic
cosmology one includes the usual
matter action $S_m$, coupled minimally to the dynamical metric,
corresponding to energy density $\rho_m$ and pressure $p_m$. Now, for
simplicity we consider the fiducial
metric
to be Minkowski\footnote{Note that this case includes the subclasses
where $f_{AB}$ can be brought to the Minkowski metric by general
coordinate transformation, as we can always choose a gauge for the
St\"{u}ckelberg fields $\phi^A$ \cite{Huang:2012pe}.}
\be
f_{AB}=\eta_{AB},
\ee
and without loss of generality we assume that the dynamical and fiducial
metrics are diagonalized simultaneously. For the dynamical metric one can
either consider for simplicity a flat Friedmann-Robertson-Walker
(FRW) form, or he can apply an open geometry. In the following two
subsections we examine these two cases separately.

\subsection{Flat universe}

We consider  a flat FRW physical metric 
of the form 
\begin{eqnarray}
\label{metricds2}
d^2 s = -N(\tau)^2 d \tau^2 +a(\tau)^2 \delta_{ij} d x^i d x^j,
\end{eqnarray}
with $N(\tau)$ the lapse function and $a(\tau)$ the scale factor, and for
simplicity for the St\"{u}ckelberg fields we choose the ansatz
\begin{eqnarray}
\label{phi0i}
  \phi^0 = b(\tau)  ,  ~~~~\phi^i =a_{ref} x^i,
\end{eqnarray}
with $a_{ref}$ a constant   coefficient.
 Although the
above specific application is only a simple subclass of the rich set of
possible scenarios, it proves to exhibit very interesting cosmological
behavior.

Variation of the total action $S+S_m$ with respect to $N$ and $a$ provides
the two Friedmann equations \cite{Huang:2012pe}:
 \begin{eqnarray}  
\label{Fr1}
3M_P^2 H^2& =& \rho_{DE}+\rho_m     ,\\
\label{Fr2}
-2 M_P^2 \dot{H}& =&\rho_{DE}+p_{DE}+\rho_m +p_m   ,
\end{eqnarray}
where we have defined the Hubble parameter $H=\dot{a}/a$, with
$\dot{a}=da/(Nd\tau)$, and in the end we set $N=1$. In the above
expressions we have defined the energy density and pressure of the
effective dark energy sector as 
\begin{eqnarray}
\label{rhode}
&&\rho_{DE} =\frac12 \dot{\psi}^2+W(\psi)+V(\psi)
\left(\frac{a_{ref}}{a}-1\right)[f_3(a) +f_1(a)] \ \ \  \ \\
\label{pde}
&&p_{DE}  =\frac12 \dot{\psi}^2-W(\psi)-
V(\psi)f_4(a)-V(\psi)\dot{b}f_1(a),
\end{eqnarray}
having also introduced the  convenient functions
\begin{eqnarray}
&&f_1(a)
=3-\frac{2a_{ref}}{a}+\alpha_3\left(3-\frac{a_{ref}}{a}\right)\left(1-
\frac{a_{ref}}{a} \right)+\alpha_4\left(1-\frac{a_{ref}} {a}\right)^2
\nonumber \\
&&f_2(a) =
1-\frac{a_{ref}}{a}+\alpha_3\left(1-\frac{a_{ref}}{a}\right)^2+
\frac{\alpha_4}{3}
\left(1-\frac{a_{ref}}{a} \right)^3
 \nonumber\\
&&f_3(a) =3-\frac{a_{ref}}{a}+\alpha_3\left(1-\frac{a_{ref}}{a}\right)
\nonumber\\
&&f_4(a) = -\left[6-\frac{6a_{ref}}{a}+\left(\frac{a_{ref}}{a}\right)^2+
\alpha_3\left(1-\frac{a_{ref}}{a}\right)\left(4-\frac{2a_{ref}}{a}
\right) +\alpha_4\left(1-\frac{a_{ref}
}{a}\right)^2\right].\ \ \ \ \ 
\label{fdefs}
\end{eqnarray}
Note that from the above expressions we observe that $a_{ref}$ plays the
role of a reference scale factor that can be arbitrary.

One can easily verify that the dark energy density and pressure satisfy
the usual evolution equation 
\begin{eqnarray}
 \dot{\rho}_{DE} +3H(\rho_{DE}+p_{DE})=0,
\end{eqnarray}
and we can also define the dark-energy equation-of-state parameter as
usual as
\begin{eqnarray}
w_{DE}\equiv \frac{p_{DE}}{\rho_{DE}}.
\end{eqnarray}
Note that in \cite{Huang:2012pe} the authors had named the aforementioned
``dark energy'' sector as ``massive gravity'' one, and the quantities
$\rho_{DE}$ and $p_{DE}$ as  $\rho_{MG}$ and $p_{MG}$. However, since in
this work we focus to late time cosmological behavior, we prefer the
above name.

Variation of the total action $S+S_m$ with respect to $\psi$ provides the
scalar-field evolution equation:
\begin{eqnarray}
\label{psievol}
&&\ddot{\psi}+3H\dot{\psi}+\frac{d W}{d \psi}
+\frac{d V}{d\psi}
\left[\left(\frac{a_{ref}}{a}-1\right)
[f_3(a)+f_1(a)]+3\dot{b}
f_2(a)\right]  =0.\ \ \ \ \ \ 
\end{eqnarray}
Furthermore, variation of $S+S_m$ with respect to
$b$ provides the constrain equation 
\begin{eqnarray}
\label{constraint}
 V(\psi)Hf_1(a)+\dot{V}(\psi)f_2(a) =0.
\end{eqnarray}
 Finally, one can also extract the matter evolution
equation
$\dot{\rho}_m +3H(\rho_m+p_m)=0$.

\subsection{Open universe}

Let us now consider an open\footnote{Similarly to usual massive
gravity, closed
FRW solutions are not possible since the fiducial Minkowski metric cannot
be foliated by closed slices  \cite{Gumrukcuoglu:2011ew,Huang:2012pe}.} FRW
physical metric of the form
\begin{eqnarray}
d^2 s = -N(\tau)^2 d \tau^2 +a(\tau)^2 \delta_{ij} d x^i d x^j
- a(\tau)^2\frac{k^2(\delta_{ij}x^idx^j)^2}{1+k^2(\delta_{ij}x^ix^j)}~,
\end{eqnarray}
with $N(\tau)$ the lapse function and $a(\tau)$ the scale factor, and $K<0$
with
$k=\sqrt{|K|}$. For
simplicity for the St\"{u}ckelberg fields we choose \cite{Huang:2012pe}:
\begin{eqnarray}
  \phi^0 = b(\tau)\sqrt{1+k^2(\delta_{ij}x^ix^j)}  ,  ~~~~\phi^i =k b(\tau)
x^i
~.
\end{eqnarray}

Variations of the action with respect to $N$ and $a$ give rise to
the following Friedmann equations
 \begin{eqnarray}
\label{Fr1b}
3M_P^2 \left(H^2-\frac{k^2  }{a^2}\right)& =& \rho_{DE}+\rho_m    
~,\\
\label{Fr2b}
-2 M_P^2\left( \dot{H}+\frac{k^2  }{a^2}\right)&
=&\rho_{DE}+p_{DE}+\rho_m +p_m,
\end{eqnarray}
where the effective dark energy density and pressure
are given by
\begin{eqnarray}
\label{rhomgb}
&&\rho_{DE} =\frac12 \dot{\psi}^2+W(\psi)+V(\psi)
\left(\frac{k b}{a}-1\right)[f_3(a) +f_1(a)] \ \ \  \ \ \\
\label{pmgb}
&&p_{DE}  =\frac12 \dot{\psi}^2-W(\psi)-
V(\psi)f_4(a)-V(\psi)\dot{b}f_1(a), 
\end{eqnarray}
but now the functions become
\begin{eqnarray}
&&f_1(a)
=3-2\frac{k b}{a}+\alpha_3\left(3-\frac{k
b}{a}\right)\left(1-
\frac{k b}{a} \right)+\alpha_4\left(1-\frac{k b} {a}\right)^2
\nonumber \\
&&f_2(a) =
1-\frac{k b}{a}+\alpha_3\left(1- \frac{k b}{a}\right)^2+
\frac{\alpha_4}{3}
\left(1 -\frac{k b}{a} \right)^3
 \nonumber\\
&&f_3(a) =3- \frac{k b}{a}+\alpha_3\left(1 -\frac{k b}{a}\right)
\nonumber\\
&&f_4(a) = -\left[6-6\frac{k b}{a}+\left(\frac{k b}{a}\right)^2+
\alpha_3\left(1- \frac{k b}{a}\right)\left(4- \frac{2k b}{a}
\right)
 +\alpha_4\left(1- \frac{k b
}{a}\right)^2\right] .\ 
\label{fdefsb}
\end{eqnarray}
These verify the usual evolution equation
\begin{eqnarray}
 \dot{\rho}_{DE} +3H(\rho_{DE}+p_{DE})=0.
\end{eqnarray}

Variation of the  action   with respect to the scalar field $\psi$ provides
its evolution equation:
\begin{eqnarray}
\label{psievolb}
&&\ddot{\psi}+3H\dot{\psi}+\frac{d W}{d \psi}
+\frac{d V}{d\psi}
\left\{\left(\frac{k b}{a}-1\right)
[f_3(a)+f_1(a)]+3\dot{b}
f_2(a)\right\}  =0 ~.\ \ \ \ \ \
\end{eqnarray}
Finally, variation  with respect to
$b$ provides the constraint equation
\begin{eqnarray}
\label{constraintb}
 V(\psi)\left(H-\frac{k }{a}\right)f_1(a)+\dot{V}(\psi)f_2(a) =0 ~.
\end{eqnarray}

\section{Cosmological behavior}
   \label{application}

The cosmological implications of extended nonlinear massive gravity, prove
to be very interesting, however, at least in its present simple but general
example, it can be radically different than the usual massive gravity. In
the following two subsections we examine the flat and open geometry
separately.

\subsection{Flat universe}
 
In the case of a flat FRW universe, the cosmological equations are 
(\ref{Fr1}),  (\ref{Fr2}) or (\ref{psievol}) and (\ref{constraint}), and
the reason that these equations lead to a different behavior comparing to
the usual massive gravity is the constraint equation (\ref{constraint}). In
order to elaborate the equations we have to consider at will $W(\psi)$ and
$V(\psi)$ and solve the equations to obtain $a(\tau)$, $\psi(\tau)$ and
$b(\tau)$, that is  the St\"{u}ckelberg scalars are suitably
reconstructed in order to correspond to a consistent solution. 

  A crucial
observation is that for $f_2(a)\neq0$
(which is
the case in general) the constraint equation (\ref{constraint}) can be
explicitly solved giving\footnote{The importance of the 
constraint equation (\ref{constraint}) was not revealed in
\cite{Huang:2012pe}, where  all the specific examples
that the authors considered were exactly those
fine-tuned
parameter choices that lead to $f_1(a)=f_2(a)=0$ and thus to a trivial
satisfaction of the constraint (\ref{constraint}).}
\begin{eqnarray}
\label{constraint2}
&&V(a) = C_0\,e^{-\int \frac{f_1}{af_2}\ud a} =
\frac{C_0}{(a-a_{ref})[
\alpha_4a_{ref}^2-(3\alpha_3+2\alpha_4)aa_{ref}+(3+3\alpha_3+\alpha_4)a^2 ]
},\ \ \ \ \ \ 
\end{eqnarray}
where we have used the definitions (\ref{fdefs}), with $C_0$ a positive
integration constant. Thus, since from the known $V(\psi)$ we can 
straightforwardly obtain $\psi(V)$ as a function of $V$,  relation
(\ref{constraint2}) eventually provides $\psi(a)$. Then one can insert the
known $\psi(a)$ into the Friedmann equation  (\ref{Fr1}) which becomes a
simple differential equation for $a(\tau)$ ($b(\tau)$ does not appear in
(\ref{Fr1})). Finally, with $a(\tau)$ known and therefore $\psi(a(\tau))$
known, one can use (\ref{psievol}) to find $\dot{b}$ as
 \begin{eqnarray}
\label{dotb}
 \dot{b}(\tau)=\frac{1}{3f_2(a(\tau))}\left\{-\frac{\ddot{\psi}
(\tau)+3H(\tau)\dot{\psi}(\tau)+\frac{dW}{d\psi}(\tau)}
{\frac{dV}{d\psi}(\tau)
}-\left(\frac{a_{ref}}{a(\tau)}-1\right)\left[f_3(a(\tau))+
f_1(a(\tau))\right]\right\},\ \ \ \ 
\end{eqnarray}
integration of which provides the St\"{u}ckelberg-scalar  function
$b(\tau)$
(note however that in the observables it is $\dot{b}$ and not $b$ that
appears).

A first observation that one can immediately make from (\ref{constraint2})
is
that in general at late
times  the graviton mass  always goes to zero,
independently of the specific $V(\psi)$ and the model parameters, that is
the evolution of $\psi$ will be such, in order for $V(\psi)$ to go to zero
(if $V(\psi)$ cannot be zero for any $\psi$ then the scenario will break
down at some scale factor, since $\psi$ would need to be complex, that
is a solution cannot be found any more). 
 This means that the present scenario of extended nonlinear massive
gravity, in a cosmological framework of a flat universe, cannot provide the
usual massive gravity, and
on the contrary it always gives the standard
gravity along with the standard
quintessence scenario \cite{quint,quint1}. Similarly, once
introduced, the scalar-field cannot be set to zero by hand, since this is
not a solution of (\ref{psievol}) and  (\ref{constraint}) (unless we also
set $V(\psi)=0$ but in this case the model coincides completely with
standard quintessence), that is $\psi$ will always have a non-trivial
dynamics.

However, although at late times the present  scenario coincides with
standard quintessence, it can have a very interesting behavior at
intermediate times. In particular, the dark energy sector is not only
dynamical, but it can easily lie at the phantom regime
\cite{Caldwell:1999ew,Dabrowski:2003jm,Chen:2008ft,Saridakis:2009pj,
Cai:2009zp,
Saridakis:2010mf}. This
can be seen
by observing $\rho_{DE}$ and $p_{DE}$ from (\ref{rhode}),(\ref{pde}),
which using the constraint equation (\ref{constraint}) give 
\begin{eqnarray}
\label{rhoplusp}
 \rho_{DE}+p_{DE}=\dot{\psi}^2-V(\psi)\left(\dot{b}-\frac{a_{ref}}{a}
\right)f_1(a).
\end{eqnarray}
So we can always find regions in the $\alpha_3$,$\alpha_4$ parameter
space, that can lead to $p_{DE}+\rho_{DE}<0$ at some stage of the
evolution (with a potential $W(\psi)$ that will not lead to large
$\dot{\psi}$), even if we require to  always have $\rho_{DE}>0$ (which does
not need to be the case in general). This null energy condition violation
is always canceled at late times, where the vanishing of the graviton mass
leads to $w_{DE}\geq-1$.

From the above discussion however one can see that despite the interesting
cosmological behavior, in the flat case there is a potential disadvantage,
namely that the
graviton  square mass, as it is given by (\ref{constraint2}), diverges and
changes sign at least for one finite scale factor independently of the
model parameters (even if we choose $\alpha_3$,$\alpha_4$ in order for the
second term in the denominator not to have roots, there is always the point
$a(\tau)=a_{ref}$)\footnote{Note that in the case where $V(\psi)$ is
imposed to be non-negative, the negativity of $V(a)$ from
(\ref{constraint2}) would demand the scalar field to be complex and thus
the model cannot have consistent solutions any more too.}. A negative
graviton square mass would make the
scenario unstable at the perturbation level and thus its application
meaningless, therefore we desire the observable universe evolution to take
place in the regime $V(\psi)\geq0$. In order to avoid a collapse of the
scenario in the future (choosing $a_{ref}$ larger than the present scale
factor) in the following we prefer to choose it suitably small in order not
to interfere with the observed thermal history of the universe
($a_{ref}\lesssim10^{-9}$ in order to be smaller than the Big Bang
nucleosynthesis scale factor). Note also that one could additionally 
``shield'' $a_{ref}$ with a cosmological bounce, case in which the
universe is always away from it \cite{Cai:2012ag}, or even choose $a_{ref}$
to be negative. However, these
considerations can only cure the problem phenomenologically, while at the
theoretical level it remains unsolved. Clearly, the scenario of a flat
universe has serious disadvantages and thus one should look for a more
general solution through its generalizations. This will be performed in the
next subsection, where the addition of
curvature makes the graviton mass square always positive. However, for
completeness we provide in the present subsection the phenomenologically
(but not theoretically) consistent flat
analysis, too.

\begin{figure}[ht]
\begin{center}
\mbox{\epsfig{figure=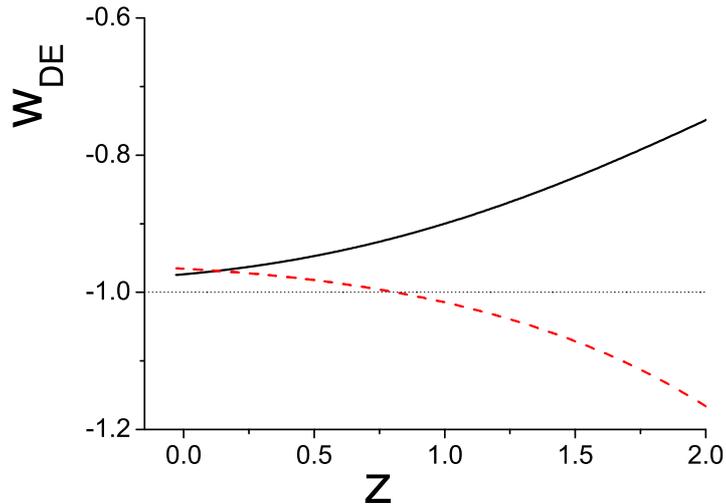,width=10.74cm,angle=0}} \caption{{\it 
Evolution
of the dark energy equation-of-state parameter $w_{DE}$
as a function of the redshift $z$, in a flat universe, in the case where
the usual scalar field
potential is $W(\psi)=W_0 e^{-\lambda_W\psi}$ and the coupling potential is
$V(\psi)=V_0 e^{-\lambda_V\psi}$. The black-solid curve corresponds to
$\alpha_3=2$, $\alpha_4=-2$, $V_0=1$, $W_0= 1.2$,
$C_0=0.001$, $\rho_{m0}=0.05$ $\lambda_V=0.7$, $\lambda_W=0.01$,
$M_P=10$, while
the
red-dashed
curve corresponds to $\alpha_3=1/3$, $\alpha_4=-2$, $V_0=10$, $W_0=5$,
$C_0=0.05$, $\rho_{m0}=0.5$, $\lambda_V=0.5$, $\lambda_W=0.02$,
$M_P=10$. All dimensional parameters are normalized in unit of $M_P$ given
that $a_0=1$, and
the dotted -1-line is depicted for convenience.
    }}
\label{flatVknown}
\end{center}
\end{figure}

In order to present the above behavior in a more transparent way,
we consider without loss of generality the graviton mass potential to be
\begin{eqnarray}
\label{Vansatz}
 V(\psi)=V_0 e^{-\lambda_V\psi},
\end{eqnarray}
and the usual scalar-field potential
\begin{eqnarray}
\label{Wansantz1}
W(\psi)=W_0 e^{-\lambda_W\psi}.
\end{eqnarray}
In this case  $\psi(a)= -\ln (V(a)/V_0)/\lambda_V$, with $V(a)$ given by
(\ref{constraint2}), and thus substitution into (\ref{Fr1}) gives a
differential equation that can be easily solved numerically to give
$a(\tau)$, while insertion into  (\ref{dotb}) provides $\dot{b}$
and therefore all the observables are known.  In
Fig.~\ref{flatVknown} we present the effective dark-energy
equation-of-state parameter $w_{DE}$
as a function of the redshift $z=a_0/a-1$ (with $a_0$ the present scale
factor set to 1), with the reference scale factor $a_{ref}$ set to
$10^{-9}$, and assuming the matter to be dust
($w_m\equiv p_m/\rho_m=0$ that is $\rho_m(a)=\rho_{m0}/a^3$, with
$\rho_{m0}$ the energy density at
present). 
The parameters $\alpha_3$,$\alpha_4$,$V_0$,$W_0$,$\lambda_V$,$\lambda_W$
are chosen at will \footnote{Note that
the graviton
mass and the usual potential are significantly downgraded by the
$\psi$-dynamics and thus they are far below  $M_P^4$ even if $V_0$ and
$W_0$ are chosen  larger than $M_P^4$.} (concerning $\alpha_3$,$\alpha_4$
we have to ensure that they lead to a positive graviton square
mass, that is especially to a positive last term in the denominator of
(\ref{constraint2})), while we fix $\rho_{m0}$ and the
integration constant $C_0$ in order for the present dark energy density
$\Omega_{DE}\equiv\rho_{DE}/(3M_P^2H^2)$ to be
$\approx0.72$ and its initial value to be $\approx0$ (concerning the
observables no more condition is needed since it is $\dot{b}$ and not
$b$ that appears in the corresponding relations, however if one desires
to obtain $b(\tau)$ too then he needs to impose an extra condition, for
instance the present $b$-value). 

As described above, at early and intermediate times the coupling potential
$V(\psi)$ is non-zero leading $w_{DE}$ to exhibit a dynamical nature, which
can lie in the quintessence regime (black-solid curve) or in the phantom
regime (red-dashed curve). Additionally, as we
said, at late times, where
the coupling $V(\psi)$ becomes zero, both sub-cases tend to their usual
quintessence limit, where the final $w_{DE}$ is determined solely from the
$W$-potential exponent $\lambda_W$ \cite{Copeland:1997et}, with the second
model experiencing the phantom-divide crossing from below to above.

In summary, as we can see the scenario at hand exhibits very interesting
cosmological behavior at early and intermediate times, with a dynamical
dark energy sector which can additionally lie in the phantom regime,
before limit towards the standard quintessence scenario. Note that despite
the phantom realization, at late times we always obtain $w_{DE}\geq-1$
since the vanishing of the graviton mass restores the null energy condition
for the effective dark energy sector, that is the universe will always
escape from the phantom regime and the Big-Rip future
\cite{Briscese:2006xu,Capozziello:2009hc} that is common to the majority of
phantom models. 

However, as we mentioned, the above flat scenario has two
significant disadvantages.
The first is that not all ansantzes for $V(\psi)$ can lead to consistent
solutions at all times, since the field $\psi$ would need to become complex
at some
scale factor, that is the theory breaks down. Secondly, the appearance
of
$a_{ref}$ in the equations leads to scale-factor regions where the
graviton mass square becomes negative, and thus the theory becomes
unstable at the perturbation level. Although one can still cure the above
problems at the phenomenological level, and move them away from the
observed universe history, clearly a generalization of the scenario is
necessary in order to completely remove these disadvantages. This is
performed if one goes beyond the flat case, as we analyze in the next
subsection.

\subsection{Open universe}

In the previous section we investigated extended massive gravity in the
case of a flat FRW universe, and we saw that the resulting cosmological
behavior can be very interesting. Although we chose the reference scale
factor $a_{ref}$ to be suitably small in order for the graviton mass
square to be always positive during the observed universe history, it is
desirable to consider a generalization of the scenario, where the potential
problem of the graviton mass square negativity will be completely absent.
This is obtained by applying extended massive gravity in a non-flat
geometry.

In the case of an open FRW universe, the cosmological equations are   
(\ref{Fr1b}),  (\ref{Fr2b}) or (\ref{psievolb}) and 
(\ref{constraintb}) (note that in this case there is no need for a
reference scale factor, since it has been absorbed inside $b(\tau)$). One
difference comparing to the flat case is that the constraint equation
(\ref{constraintb}) cannot be solved analytically and thus it has to be
considered along the other cosmological equations. Although this brings an
additional mathematical complexity, it offers a great physical
advantage, since the constraint satisfaction can be obtained by
significantly larger solution subclasses, and therefore one can  always,
and in general, find solutions where the graviton mass square is always
positive and finite. Similarly to the flat case, in the following
 we   consider at will the usual scalar field potential $W(\psi)$ and the
coupling potential $V(\psi)$ and we solve the
equations to obtain $a(\tau)$,  $\psi(\tau)$  and $b(\tau)$.

Let us consider known forms for  $W(\psi)$
and   $V(\psi)$. Due to the constraint dependence on
$b(\tau)$ it cannot be solved alone, and thus one needs to solve the
whole system of equations simultaneously. Since this is not
analytically possible we proceed to a numerical elaboration of a specific
example. In particular, we first solve algebraically (and  analytically
if it is possible) the constraint (\ref{constraintb}) in order to extract
$b(\tau)$ as a function of
$a(\tau)$,$\dot{a}(\tau)$,$\psi(\tau)$,$\dot{\psi}(\tau)$ and then
substituting the resulting (quite complicated) expression into
(\ref{Fr1b}),(\ref{Fr2b}) we obtain two differential equations for
$a(\tau)$ and $\psi(\tau)$ that do not depend on $b(\tau)$, which can be
numerically solved. Note that contrary to the flat case $V(\psi)$ does not
need to be able to become zero at some $\psi$ in order for the equations
to be solvable, however for phenomenological reasons we do consider it to
be able to reach zero or very small values chosen at will and in
agreement with experimental bounds (thus in this case one can
re-obtain the usual non-flat massive gravity, where the graviton mass is
very small but non-zero). 
\begin{figure}[ht]
\begin{center}
\mbox{\epsfig{figure=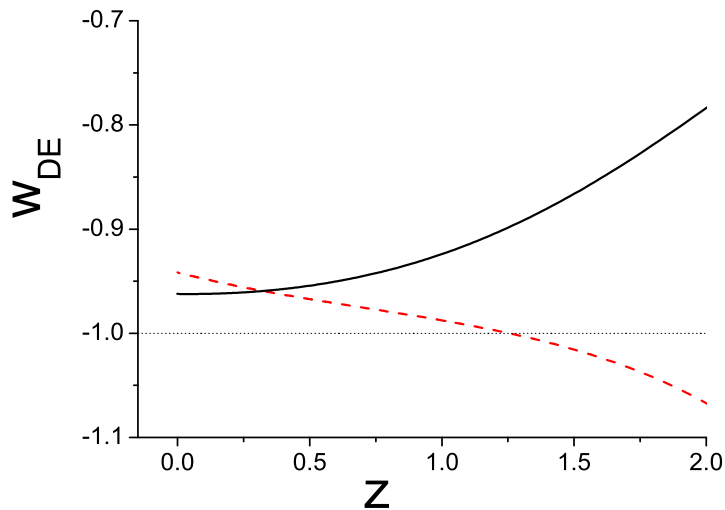,width=10.74cm,angle=0}} \caption{{\it
Evolution
of the dark energy equation-of-state parameter $w_{DE}$
as a function of the redshift $z$, in an open universe,
in the case where the usual scalar field
potential is $W(\psi)=W_0 e^{-\lambda_W\psi}$ and the coupling potential is
$V(\psi)=V_0 e^{-\lambda_V\psi}$. The black-solid curve corresponds to
$\alpha_3=1$, $\alpha_4=1$, $k=0.02$, $V_0=0.6$, $W_0= 0.4$, 
$\rho_{m0}=0.09$,
$\lambda_V=6$, $\lambda_W=0.4$,
$M_P=1$ (with $\dot{a}_0=0.3$, $\psi_0=1.7$,
$\dot{\psi_0}=0.09$) while
the
red-dashed
curve corresponds to $\alpha_3=2$, $\alpha_4=2$, $k=0.3$, $V_0=2.4$, $W_0=
4$, $\rho_{m0}=0.04$,
$\lambda_V=5.5$, $\lambda_W=0.6$,
$M_P=1$  (with  $\dot{a}_0=0.9$, $\psi_0=1.5$,
$\dot{\psi_0}=0.33$). All dimensional parameters are normalized in unit of
$M_P$ given that $a_0=1$, and
the dotted -1-line is depicted for convenience. }}
\label{nonflatV}
\end{center}
\end{figure}

We choose both $V(\psi)$  and
 $W(\psi)$ to have the exponential forms (\ref{Vansatz}) and
(\ref{Wansantz1}), namely  $ V(\psi)=V_0 e^{-\lambda_V\psi}$ and 
$W(\psi)=W_0 e^{-\lambda_W\psi}$ respectively, although we could still add
a  constant in $V(\psi)$, suitably small in order to be consistent with
experimental bounds. We evolve the system
numerically, using the redshift $z=a_0/a-1$ as
the independent variable (with $a_0=1$ the present scale factor),  and
assuming dust matter ($\rho_m(a)=\rho_{m0}/a^3$, with
$\rho_{m0}$ the present energy density). The parameters
$\alpha_3$,$\alpha_4$,$V_0$,$\lambda_V$,$W_0$,$\lambda_W$  are chosen at
will, while we fix $k$ in order for the present curvature density parameter
($\Omega_k=k^2/(a^2H^2)$) to be $0.01$, and we fix the present values
$\rho_{m0}$, $\psi_0$, $\dot{\psi}_0$ and $\dot{a}_0$ in order for the
present dark energy density
$\Omega_{DE}\equiv\rho_{DE}/(3M_P^2H^2)$ to be
$\approx0.72$, its initial value to be $\approx0$, and the present
dark-energy equation-of-state parameter to be between $-0.9$ and $-1$ in
agreement with observations. 

In fig.~\ref{nonflatV} we present $w_{DE}$ as a function of $z$, for two
choices of the parameters. As we observe, at early and intermediate times
the coupling potential $V(\psi)$ is non-zero leading $w_{DE}$ to exhibit a
dynamical nature, which can lie in the quintessence regime (black-solid
curve) or in the phantom regime (red-dashed curve), and it can cross the
phantom divide from
below to above, before asymptotically limit towards the usual quintessence
scenario. This behavior is similar to the flat universe,
however as we
mentioned, in the present case the graviton mass square is always finite
and positive, independently of the specific solution.

 \section{Discussion}
\label{Discussion}

In this work we investigated the cosmological evolution in a universe
governed by the extended, varying-mass, nonlinear massive gravity. Even
for simple ansatzes the scenario proves to have a very interesting
behavior, comparing with standard massive gravity.

The first result is that the dynamics in cosmological frameworks  
can lead the varying graviton mass to zero at late times, both in flat and
open geometry (in the open case one can also obtain at will a non-zero
but suitably small value if he correspondingly choose the coupling
potential),
and thus the theory possesses as a limit the standard
quintessence paradigm. This is a great advantage of the present
construction, since it offers a natural explanation of the tiny and
hugely-constrained graviton mass that arises from current observations. The
graviton mass does not have to be tuned to an amazingly small number, as it
is the case in standard massive gravity, but it is the dynamics that
can lead
it asymptotically to zero. Additionally, although in the simple flat case
one may face the problem of a divergent or negative graviton mass square,
which should be then shielded by a cosmological bounce, in the non-flat
scenario the graviton mass square is always finite and positive,
independently of the specific solution.

Despite the vanishing of the graviton mass at late times, and the limit of
the scenario towards standard quintessence, at early and intermediate ones
it can lead to very interesting behavior. In particular, it can give rise
to an effective dark energy sector of a dynamical nature, which can also
lie in the phantom regime. The violation of the null energy condition for
the effective dark energy sector at intermediate times arises naturally for
suitable (not fine-tuned) regions in the Lagrangian parameters, and it is
always canceled at late times due to the vanishing of the graviton mass.
These features are in agreement with observations and they
offer an explanation for the dynamical evolution of the dark-energy
equation-of-state parameter, for its relaxation close or at the
cosmological constant value, and also for the indicated possibility to have
crossed the phantom divide. Moreover, even if it enters the phantom
regime, the scenario at hand always returns naturally to the quintessence
one, offering a solution to the Big-Rip fate of the   standard phantom
scenarios. The complete investigation of the possible late-time behaviors
is performed in \cite{Leon:2013qh}, through a detailed dynamical
analysis.

We mention here that although we performed the above analysis with the
fiducial metric to be Minkowksi, and with specific ansantzes for the
potentials and the St\"{u}ckelberg-scalars,  qualitatively the
obtained behavior is not a result of them, but it arises from
the deeper structure of the theory, namely from the scalar-field coupling
to the graviton potential. Thus, we do not expect the
results to change in more general cases, unless one fine-tunes the theory.

In the above analysis we remained at the background level, as a first
approach to the examination of the properties of the theory. Obviously, a
crucial issue is the complete investigation of the perturbations, in order
to see whether the scenario at hand suffers from instabilities. Although
one could be based on similar studies of usual massive gravity
\cite{Gumrukcuoglu:2011zh,Crisostomi:2012db,DeFelice:2012mx,D'Amico:2012pi,
Fasiello:2012rw}, and see that the generalized Higuchi bound is satisfied, 
we mention that since a cosmic scalar is introduced to drive the graviton
mass varying along background
evolution, the stability issue arisen from this scalar field ought to be
taken into account in a global analysis. Such a complete perturbation
analysis of the extended nonlinear massive gravity lies beyond the scope of
the present work and it is left for future investigation.
 
In conclusion, the extended, varying-mass, nonlinear massive gravity leads
to very interesting cosmological behavior at early and intermediate times,
while it   limits towards the standard quintessence scenario, where
the graviton is massless and the extra scalar is only minimally coupled to
gravity. Strictly speaking, although the motivation of massive gravity is
to obtain an IR modification, its varying-mass extension  in cosmological
frameworks leads rather to early and intermediate times modification, and
thus to a UV modification instead.

\vskip .2in \noindent {\large{{\bf {Acknowledgments}}}}

We wish to thank Y-F. Cai,   A.~Gumrukcuoglu, Q-G. Huang,  T.
Koivisto, C.~Lin, S.~Mukohyama, Y.-S, Piao, J. Stokes,               
N. Tamanini and S.-Y. Zhou, for useful comments.
The
research project is
implemented within the
framework of the Action «Supporting Postdoctoral Researchers» of the
Operational Program ``Education and Lifelong Learning'' (Action’s
Beneficiary: General Secretariat for Research and Technology), and is
co-financed by the European Social Fund (ESF) and the Greek State.\\

\providecommand{\href}[2]{#2}

\begingroup

\raggedright

\endgroup

\end{document}